\documentclass{article}
\usepackage[utf8]{inputenc}
\usepackage{hyperref}
\usepackage[margin=1in]{geometry}
\usepackage{makecell}
\usepackage[onehalfspacing]{setspace}
\makeatletter
\renewcommand{\@biblabel}[1]{}
\makeatother

\title{Community Notes: Crowd Participation and Dependence on Professional Fact-Checking Across Languages}
\author{
Elizabeth Stewart\footnote{Corresponding author: elizabeth.stewart@canterbury.ac.nz}\\
University of Canterbury
\and
Suryash Greenwold\\
University of Canterbury
\and
Timotius Marselo\\
University of Canterbury}
\date{\makecell{PREPRINT\\December 2025}}

\begin{document}

\maketitle

\begin{abstract}
Crowd-sourced fact-checking provides social media platforms with a promising method of managing misinformation at scale. However, the success of fact-checking programs like X's Community Notes requires the participation of a critical mass of note-writers who have the time and epistemic resources necessary to write and rate high-quality notes. As X's Community Notes program was first established in English-speaking countries, much academic research has focused on English-language notes or notes writ large. Relatively little research has investigated how different linguistic communities utilise Community Notes. Thus, it is unclear whether Community Notes or similar crowd-sourced fact-checking initiatives represent a viable alternative to social media platforms' partnerships with professional fact-checking organisations across linguistic contexts. This research identifies how different linguistic communities participate in volunteer fact-checking efforts on X's Community Notes program and addresses volunteers' reliance on professional fact-checking resources differs across languages. We find that while the Community Notes program has had strong uptake in some linguistic communities, the program has failed to catch on in others. Additionally, we confirm findings that notes that cite professional fact-checkers are considered more helpful, but show that reliance on professional fact-checking overall remains minimal.
\end{abstract}

\section*{Keywords}
Content Moderation, Fact-Checking, Social Media, Misinformation, Crowd-sourcing\\
\\
Word Count: 6,799

\newpage

\section{Introduction}
Social media platforms’ content moderation strategies continue to evolve in response to a persistent flood of misinformation online. Misinformation differs from other forms of problematic online content, such as harassment or trolling, insofar as its identification often requires domain expertise or access to veridical information about the world. Given that misinformation contingently depends on the nature of the world and how events unfold in real time, it is incredibly difficult to use automated tools to identify novel instances of misinformation. As such, addressing misinformation online involves substantial amounts of human labor. 

Platforms have pursued two primary strategies for sourcing this human labor: to contract with third-party fact-checkers and to use community-generated fact-checks. Both strategies have important limitations \cite{augensteinetal2025,pilarskietal2024,godeletal2022}. Professional fact-checkers provide in-depth analyses of complex events, drawing on their own journalistic expertise and resources, including advanced reverse-image search tools and the social capital to seek statements from involved parties, that are often not available to volunteer fact-checkers. However, professional fact-checking struggles to keep up with the sheer volume of misinformation shared online. Thus, crowd-sourcing fact-checking may offer an easily scalable response to misinformation, although such fact-checks may individually provide less information than a professionally produced fact-check would. Thus, Augenstein et. al. (2025) argue that platforms would likely benefit from fostering collaboration between volunteer and professional fact-checkers.

Additionally, there is evidence that a dependent relationship already exists between volunteer fact-checkers and their professional counterparts. As X/Twitter has the most well-established crowd-sourced fact-checking program and makes its data freely available, most studies on crowd-sourced fact-checking are drawn from X’s Community Notes program (formerly Birdwatch). Borenstein et al. (2025) demonstrate that at least 5\% of all English community notes contain an external link to professional fact-checkers. Of the notes rated as ``Helpful'', and which are therefore published, that number increases to 7\%. Solovev and Pröllochs (2025) did not filter out non-English notes, finding that while only 3.34\% of links were to third-party fact-checkers, notes that linked to these were perceived as particularly helpful \cite{solovevandprollochs2025}. Similarly, a report produced by the Fundación Maldita.es found that while only 8.3\% of proposed notes reach enough agreement for publication, that number increases to 12\% when the notes cite a professional fact-check \cite{maldita2025}. 

Understanding whether there are any dependent relationships between crowd-sourced and professional fact-checking is particularly important in light of Meta's decision in January 2025 to end its partnership with third-party fact-checkers in favor of a Community Notes model of crowd-sourcing fact-checks. While this change is initially limited to the US, Meta may roll out this model worldwide. Doing so may have knock-on effects for the professional fact-checking industry as many have previously relied on Meta's funding. This could be a particularly important problem for linguistic communities that have fewer professional fact-checking resources available in their language. For linguistic communities with only one professional fact-checking organisation, its closure would leave potential volunteer fact-checkers, and their audiences, without access to these informational resources.

If notes that cite a professional fact-checking organisation are rated as more helpful, as previous research suggests, then leaving some linguistic communities with no access to these would place them in a worse position to write and receive helpful notes. There is existing evidence that platforms do not equitably dedicate moderation resources across language groups \cite{globalwitness2023}. If Community Notes writers depend on professional fact-checkers to produce high-quality notes, but those fact checkers depend on social media platforms' funding, then this inequity could be exacerbated. In order to identify potential inequities in access to quality informational sources and how that impacts the efficacy of Ccommunity Notes-style programs in tackling misinformation, it is vital to understand the extent to which Community Notes contributors depend on professional fact-checkers. To address this question, this project focuses on the following research questions:\\ \textbf{RQ1: Does participation in volunteer fact-checking differ across linguistic communities?}\\
 \textbf{RQ2: Does reliance on professional fact-checking organisations differ by linguistic community?}

\section{Background}
\subsection{Professional Fact-Checking}
Historically, professional fact-checking developed in response to political misinformation, but during the 2010's expanded to address misinformation related to a wide range of topics, including health, science, ongoing conflicts, and celebrity gossip. While fact-checking grew out of the broader journalism profession, over the past two decades the field has developed its own professional norms and practices \cite{gravesandamazeen2019}. The founding of the International Fact-Checking Network (IFCN) in 2015 exemplifies the increased professionalisation of the field, as it laid out clear standards for fact-checkers and initiated a certification process to enforce them. These standards include commitments to fairness, transparency of informational sources and funding sources, rigorous standards for and transparency of fact-checking methodology, and an open and honest corrections policy \cite{ifcn2025}. In order to gain IFCN certification, fact-checking organisations must submit an application that is assessed by independent assessors who check for compliance with the above standards. 

In December 2016, Meta began partnering with professional fact-checking organisations in order to address misinformation on Facebook \cite{meta2023}. Meta requires that its professional fact-checking partners have either IFCN verification or membership in the European Fact-Checking Standards Network \cite{meta2025}. This ensures that its fact-checking partners meet certain informational standards. The partnership involves fact-checkers assessing claims that are identified as potential misinformation via Meta's proprietary algorithms, other users, or the fact-checkers themselves. Based on the fact-checkers' verdicts, posts deemed ``False'', ``Altered'', or ``Partly False'' receive reduced distribution on Facebook, with fact-checks appended to these posts.  

Meta's partnership represents a significant source of funding for many IFCN signatories. In 2024, Meta represented one of the two largest revenue sources for fact-checking organisations, the other being grants \cite{ifcn2024}. Following its decision to gradually end its third-party fact-checking program, the loss of Meta's contracts poses a significant challenge for the financial sustainability of some professional fact-checking organisations, with a small handful (7.8\%) almost entirely dependent on Meta's funding. In the IFCN's 2024 \textit{State of the Fact-Checkers Report} 30\% of fact-checking organisations expected that in light of Meta's decision to end its partnership they would need to reduce fact-checking output, 29\% anticipate cutting jobs, and 8\% may shut down entirely. Only 26.7\% of IFCN signatories did not receive any funding from Meta.

While the aim of social media platforms' established partnerships with professional fact-checking organisations is to tackle online misinformation and research shows that fact-checking does reduce false beliefs \cite{walteretal2019,porterandwood2021}, the practice has been long been subject to criticisms of bias and unfair censorship \cite{uscinskiandbutler2013}. Responding to these criticisms in January 2025, Mark Zuckerberg announced the end of Meta's partnership with professional fact-checkers in the United States, claiming that fact-checkers were ``too politically biased and have destroyed more trust than they've created'' \cite{zuckerberg2025}. Shifting to a crowd-sourced Community Notes model of fact-checking creates a buffer against these criticisms, as fact-checking becomes a democratised practice, open to anyone. 

\subsection{Crowd-Sourced Knowledge Production}
There is precedence for crowd-sourcing knowledge production in online contexts. Twitter, now X, began its crowd-sourced fact-checking program in 2021. However, long pre-dating this, Wikipedia has relied on the ``wisdom of the crowd'' since its creation in 2001, becoming the largest and most widely used encyclopaedia worldwide \cite{mcdowellandvetter2022}. Several studies evaluating the accuracy of Wikipedia show that, while not perfect, it is generally comparable to other encyclopaedias and reference texts \cite{giles2005,kraenbringetal2014,bragues2009,brown2011}. Wikipedia's success depends on large numbers of volunteer contributors who hold each other accountable for following policies of neutrality and verifiability. Yet, many of the studies documenting Wikipedia's accuracy also note that it contains glaring omissions, both in terms of details within existing articles and the lack of articles on certain topics. 

For example, Wikipedia has long been known for its gender and cultural bias, with contributors heavily skewing white, Western, educated, and male \cite{mcdowellandvetter2022,fordandwajcman2017}. The lack of diverse representation has had a number of consequences, most famously, perhaps, is it's gender gap \cite{fordandwajcman2017}. As of the end of 2024, only 20\% of biographies on Wikipedia are of women \cite{wirwiki}. Additionally, gender-based harassment remains an issue in the editor community and the gatekeeping of newcomers is a barrier that often discourages newer, more diverse, contributors \cite{mcdowellandvetter2022,fordandwajcman2017}.  

In addition to gender bias, Wikipedia's emphasis on verifiability via written text marginalises orally transmitted indigenous knowledge \cite{mcdowellandvetter2022}. As an encyclopaedia, Wikipedia has a policy that prohibits original research; all information added to an article must originate from a reputable written source that is, ideally, linkable and accessible to anyone on the internet. While this enables users to easily verify content, it comes with the consequence that certain forms of knowledge, such as first-hand experience or oral testimonies, cannot be directly included.

While Wikipedia is still wrestling with lack of diversity, especially with respect to gender, it has worked toward linguistic diversity. There are 342 different language editions currently available as of December 2025. The English language edition remains the largest and most used, representing 10.8\% of all articles written \cite{wikieditions}. However, the top ten largest editions contain nearly half of all written articles, with the remaining half distributed across over 330 languages. Furthermore, non-English articles tend to be shorter, less detailed and include fewer references \cite{royetal2022}. Thus, while many languages are represented on Wikipedia, there are clear disparities in the quantity and quality of information made available via the platform. 

Accordingly, while Wikipedia demonstrates the viability of crowd-sourced knowledge production, it also highlights some of the limitations. Crowd-sourcing requires an online community of diverse volunteers with time and willingness to contribute their expertise for free. This requires barriers for entry that are flexible enough to welcome newcomers, while also discouraging and removing hostile or bad-faith actors. Additionally, crowd-sourcing requires reliable written sources that volunteers, who may lack domain expertise, can access and understand. Without such diversity and informational resources, the crowd-sourced knowledge base risks glossing over important topics or failing to include them entirely. 

\subsection{Bridging Algorithms}
In an attempt to mitigate potential biases, both X and Meta use \textit{bridging algorithms} to determine which notes get shown publicly on the platform \cite{wojcik2022,zuckerberg2025}. Bridging algorithms are designed to reward content that appeals to ideologically diverse audiences \cite{ovadya2022}. This is in contrast to engagement-based recommendation algorithms, which reward content that elicits strong responses, both positive and negative. Such engagement-based algorithms tend to reward politically divisive content, as this evokes strong responses. Bridging algorithms rely on users' historical patterns of engagement to select for content that would elicit positive responses from users who have disagreed in the past. Put simply, bridging algorithms prioritise content that appeals to people on both sides of a political disagreement. 

 Meta's Community Notes program is modelled after X's and, as X has made their algorithms and Community Notes data publicly available, this discussion will focus on X, unless otherwise noted. 
 When a note on X is written, rather than being automatically appended to a post, users enrolled in the Community Notes program vote on whether they find the note helpful or not. Notes that secure enough agreement are then appended to the corresponding public post. Thus, notes can have one of three statuses: ``Helpful'', ``Not Helpful', and ``Needs More Ratings''. ``Helpful'' notes are those that a sufficient number of ideologically different contributors rate as helpful. Once rated as helpful, these notes are publicly shown beneath the corresponding X post. ``Not Helpful'' notes are those that a sufficient number of ideologically diverse contributors rate as unhelpful. These notes do not appear to the general public. Notes with ``Needs More Ratings'' lack sufficient ideologically diverse agreement and do not appear to the public unless they receive more ratings. A note's status is typically ``locked'' after two weeks (but see discussion on different models below for exceptions), at which point the note's status will no longer change regardless of any further votes. Importantly, whether a note is deemed Helpful is \textit{not} the result of a majority vote as such a system would encourage biased note writing. Instead, a note's helpfulness is determined by bridging algorithms, which take as inputs the \textit{number} of votes as well as the \textit{voting history} of raters. This ensures that notes that have a "Helpful" status and appear to the public will appeal to a diverse audience, hopefully reducing partisan bias \cite{xnotemodels2025}. Unfortunately, mirroring the limitations of Wikipedia, bridging algorithms do not address biases due to omission. If there are no users volunteering their time to write notes on a given topic or in a given language, then that content will not be subject to fact-checking.

Anyone can theoretically join X's Community Notes. After being admitted, users can rate others' notes as Helpful or Not Helpful. When rating, X provides a multiple choice response so that raters can indicate the reason or reasons they found it ``Helpful'', ``Somewhat Helpful'', or ``Not Helpful''. For example, a note rater can indicate that they found the note Helpful because it cites high quality sources, is easy to understand, directly addresses the post's claim, provides important context and/or uses neutral or unbiased language. When a participant is initially added to the program, they only have the ability to rate notes. After they receive a ``Rating Score'' of 5, meaning they've helped five notes achieve a status of Helpful or Not Helpful, participants unlock the ability to write notes. A note writer can unlock further abilities, including the ability to write notes on images and videos and the ability to see when X users request a Community Note, if they achieve a status as a ``Top Writer''. Top Writers must have a ``Writing Impact'' score of ten or more and have at least 4\% of all their notes rated helpful. Writing impact increases when their notes earn a status of ``Helpful'' and decrease when a note earns the status of ``Not Helpful''. This hierarchical gatekeeping aims to prevent trolls or other bad faith actors from participating.

There is emerging research that suggests that, at least in some domains, Community Notes that are rated as Helpful also tend to be accurate \cite{allenetal2024}. However, others have criticised the bridging algorithms that X uses because they result in a relatively small number of posts published relative to the number of posts written. Previous research suggests that only approximately 10\% of notes get published \cite{maldita2025,borensteinetal2025}. One report that focused on election misinformation showed that 74\% of accurate notes debunking election misinformation failed to secure enough agreement for publication \cite{ccdh2024}. Thus, bridging algorithms prioritise topics around which users can find consensus, but may neglect those which are verifiably true or false, but contentious \cite{prollochs2022}. Additionally, the voting process required for bridging algorithms takes time, and some research suggests that the process is too slow to reduce the spread of misinformation in the early stages when it is most viral \cite{chuaietal2024}.
 
 X has continuously updated its algorithms in order to mitigate attempts at coordinated misinformation campaigns and also to accommodate newer entrants to the Community Notes program. When Community Notes (then called ``Birdwatch'') began in 2021, it was only available in the United States. The program progressively expanded until 2024, when it was made available everywhere X is supported. X employs five different algorithms across different geographical areas depending on how long those areas have been part of Community Notes and how active a particular linguistic community is \cite{xnotemodels2025}. Thus, in areas where Community Notes is newly deployed and in non-English speaking communities, X uses different algorithms in order to account for fewer numbers of participants. These are designed to make it easier for a note to move from a status of ``Needs More Ratings'' to ``Helpful''. This feature is important for the current study as we are interested in how Community Notes functions across linguistic communities of different sizes and across geographical locations that have been added more recently to the program. 
 
X's \textit{Core} model determines the status for notes written in geographical areas where Community Notes is well established, such as the United States. The \textit{Expansion} and \textit{ExpansionPlus} models are used for areas where Community Notes has recently launched and is not yet well-established. The \textit{Group} model improves note ranking in non-English speaking communities. Finally, the \textit{CoreWithTopics} model is applied to a predefined set of topics, enabling X to maintain a higher standard of helpfulness across viewpoints for those topics. After two weeks, a note's status is ``locked'' and can no longer be updated, unless it was ranked by the \textit{ExpansionPlus} model or the \textit{TopicModel}. These latter models are still under development and so notes are not locked as the models and their associated rankings may be updated to improve quality. If these algorithms successfully account for fewer numbers of participants across diverse geographic and linguistic contexts, then we should see similar rates of ``Helpful" and ``Needs More Ratings" statuses across linguistic communities, regardless of size and how long Community Notes has been available in a particular location.
 
 In what follows, we analyze how often notes are written in different languages and the extent to which notes reach a verdict of ``Helpful'' or ``Not Helpful'' versus ``Needs More Ratings''. Additionally we analyse the extent to which different languages rely on professional fact-checking sources.
 
 \section{Analysis}
 \subsection{RQ1: Does the rate of note publication differ across languages?}
\subsubsection{Methodology}
To assess global participation in X's Community Notes program, we accessed X's publicly available Community Notes data. X makes a range of information publicly available, including the text of the note and any links added for supporting evidence, the unique ID of the person who wrote the note, the history of the note's status, and the reason the note writer found the post problematic. For our research, we used the ID of the note writer, the text of the note and its current status (``Helpful'', ``Not Helpful'', ``Needs More Ratings''). Importantly, X does not provide access to the public posts that correspond to the notes. Thus, we did not attempt to analyse the topics or content that the notes targeted.

We selected all notes written between September 2024 and June 2025. We did not include notes written prior to September 2024 because the Community Notes program wasn't yet available worldwide. We also did not include notes written after June 2025 because, at the time of analysis (July 2025) we wanted a two week buffer so that notes would have time to reach a locked status. Looking at this time frame, then, captures worldwide participation in the Community Notes program. This dataset included 671,358 community notes. 

Using fastText language identification \cite{graveetal2018}, we classified notes according to language. This provided information as to how many total notes are written in each language. We then used X's data on the current status of notes to determine how often notes across languages reach a status of ``Helpful'' and are thereby made publicly visible, versus how often notes are either rated ``Not Helpful'' or in need of more ratings. Additionally, we assessed how many users contributed notes in each language. We did not count the total number of note writers, as some note writers may contribute notes in different languages. Instead, for each language, we assessed how many users wrote at least one note in that language.

\subsubsection{Results}
While notes were written in 79 different languages, the majority of these were written in English (413,950). The top five most common languages represented in the dataset (English, Spanish, Japanese, Portuguese and French) accounted for over 90\% of the notes written. Unsurprisingly, we found that English had the most unique writers by far. Across the top fifteen languages, note writers contributed about three and a half notes on average. Japan was an outlier with note writers contributing an average of over five notes. The results for the 15 languages with more than 1,000 notes are included in Table 1. 
\begin{table}[h!]
\begin{center}
\begin{tabular}{|c|c|c|c|c|c|c|}
\hline
Language & Date Added & \makecell{Number\\of Contributors} & \makecell{Number\\ of Notes} & \makecell{Notes Rated\\ ``Helpful''} & \makecell{Notes Rated \\``Not Helpful'' }& \makecell{Notes Needing \\More Ratings}\\
\hline
English & Oct. 2022 & 104,852 & 413,950 & 7.81\% & 4.13\% & 88.06\% \\
\hline
Spanish & April 2023 & 18,802 & 62,753 & 9.15\% & 3.22\% & 87.63\% \\ 
\hline
Japanese & March 2023 & 11,494 & 58,776 & 11.55\% & 3.23\% & 85.22\% \\
\hline
Portuguese & March 2023 & 9382 & 42,140  & 12.95\% & 3.50\% & 83.55\% \\
\hline
French & Dec. 2022 & 9237 & 34,576 & 11.23\% & 4.10\% & 84.67\% \\
\hline
German & June 2023 & 4,440 & 14,891 & 7.14\% & 3.04\% & 89.82\% \\
\hline
Turkish & July 2023 & 3,819 &14,621 & 16.50\% & 3.05\% & 80.45\% \\
\hline
Hebrew & Nov. 2023 &1,361 & 5,816 & 12.00\% & 4.37\% & 83.63\% \\
\hline
Polish & July 2023 & 1,357 & 5,523 & 5.83\% & 1.53\% & 92.64\% \\
\hline
Italian & June 2023 &1,199 & 3,224 & 7.97\% & 3.26\% & 88.77\% \\
\hline
Arabic & Nov. 2023 & 896 & 2,269 & 7.27\% & 1.14\% & 91.32\% \\
\hline
Dutch & June 2023 & 782 & 2,013  & 6.86\% & 3.23\% & 89.91\% \\
\hline
Serbian & July 2023 & 494 &1,818  & 24.26\% & 0.99\% & 74.75\% \\
\hline
Indonesian & Nov. 2023 & 429 & 1,148 & 12.80\% & 1.13\% & 86.07\% \\
\hline
Persian & Sep. 2024 & 348 & 1,019 & 10.20\%& 2.06\% & 87.74\% \\
\hline
\end{tabular}
\caption{Languages with more than 1,000 Community Notes. The ``Date Added'' column details the date when Community Notes was first made available in a country where that language is an official or recognized language.}
\label{table:1}
\end{center}
\end{table}

Thirty-four languages had 10 or fewer notes, 15 of which had only one note written in that language. Of those languages with 10 notes or fewer, two were constructed languages (Esperanto and Ido). These results demonstrate disparate rates of participation in the Community Notes program across linguistic communities. This difference is likely due to a range of factors, including (1) how long Community Notes has been available in the geographic areas where that language is primarily spoken, (2) the overall number of speakers, and (3) how many of those speakers are active on X.

 As can be seen in Table 1, the top five languages represented are all spoken in countries where Community Notes has been operating the longest. According to the Community Notes account on X, after a period of pilot testing, the feature was made public in the US in October 2022, followed by Canada in December 2022, with the primarily English-speaking countries of New Zealand, Australia, Ireland and the UK added in January of 2023 \cite{cntweet1}. Brazil and Japan were added in March 2023, with Portugal, Spain, Mexico and a number of other Latin American and South American countries added over April and May of that year. While Canada was added late 2022, other French speaking countries followed in June 2023. 

Of the 15 languages with only one note, one is a fabricated language (Ido) and five (Irish, Yiddish, Low Saxon, Occitan and Asturian) are on the UNESCO list of vulnerable or endangered languages, indicting relatively few overall numbers of speakers \cite{moseley2010}. Luxembourgish, while not listed as vulnerable, is in competition with French which is notably one of the top five most represented languages in Community Notes. Additionally, seven of these languages are primarily spoken in countries that were not added to the Community Notes program until it was made available worldwide in September 2024 (Dhihevi, Macedonian, Minangkabau, Armenian, Kurdish, Somali and Sorani).

We did not have access to the tweets posted over this time period and as such it is difficult to know how active different linguistic communities are on X, broadly speaking. However, some recent research has been done to identify which geographical regions are most active on X \cite{sohailetal2021,sohailetal2025}. This is a difficult task as, at the time of writing, the geographical origin of tweets wasn't available directly from X and thus had to be inferred from information about and provided by platform users. In their analysis, they were able to infer the origin of roughly 75\% of the tweets in their dataset. From this, they identified that the countries that contribute the most posts to X are (in order of contribution): the United States, Brazil, United Kingdom, Spain, France, Argentina, India, Mexico, Canada, and Nigeria \cite{sohailetal2021}. Insofar as some languages are geographically constrained, this gives some indication as to which linguistic communities are likely most active on X. It is thus not surprising that of the five top languages in Community Notes, all but Japanese, are spoken in these highly active areas. India stands as an exception to this, with less than 700 notes written in languages spoken in India (391 notes written in Hindi, 111 written in Urdu, 96 written in Tamil, 34 in Malayalam, 25 in Marathi, 6 in Kannada, 4 in Telugu), despite having an active presence on X.  

The discrepancy between India's active presence on X and relatively few notes may be due to it being a recent addition to the Community Notes program. Whereas X deployed the Community Notes program in the other highly active geographical areas early on, India wasn't added until April 2024, only 5 months before the earliest data in our dataset. Conversely, Japan was added to the program relatively early in March 2023 and, despite not being in the group of most active tweet-writing countries, Japanese is the third largest contributing language to Community Notes. However, Japan's rate of Community Notes participation may have less to due with how long the program has been available in Japan and more due to the fact that its note writers are particularly active, contributing about 5 notes on average (as compared to 3.4 on average for the other 14 top languages). 

The number of notes written, however, isn't a sufficient indication of the quality of those notes. Effective community moderation requires not only that notes are written, but that these notes are helpful. We found a lot of variation in how often notes were rated as Helpful or Not Helpful across languages. Of the fifteen languages with more than 1,000 notes, Polish language notes were least likely to reach a status of Helpful (5.83\%), followed by Dutch (6.86\%). Conversely, Serbian had the highest rate of Helpful notes (24.26\%), followed by Turkish (16.5\%). Serbian also had the lowest proportion of notes rated as Not Helpful, with less than 1\% of written notes rated as Not Helpful. Hebrew had the highest rate of notes rated as Not Helpful (4.37\%), followed by English (4.13\%). 

Unsurprisingly, there is a general trend that languages with fewer notes have lower rates of notes reaching a Helpful status. In languages with over 1,000 notes, the average rate of notes reaching a determinate status of Helpful or Not Helpful is 13.72\% (10.9\% are rated Helpful), whereas for languages with between 100 and 999 notes, this is only 9.95\% (7.34\% are rates Helpful). Among the languages least represented, it is unsurprising that few of these notes reached a determinate status of Helpful or Not Helpful. For the 48 languages with fewer than 100 notes, 42 had no notes rated as Helpful. Interestingly though, when it came to rating notes as Not Helpful, 17 of these languages had one or more notes rated as Not Helpful. This is in contrast to languages with more than 100 notes, which almost universally had a higher proportion of notes rated as Helpful than Not Helpful.

\subsection{RQ2: Does reliance on professional fact-checking organisations differ by linguistic community?}

\subsubsection{Methodology}
To assess the extent to which different linguistic communities rely on professional fact-checkers, we first checked the overall reliance on IFCN signatories across all notes in our dataset (those written between September 2024 and June 2025). We identified all IFCN signatories, both those currently active and those in the process of renewing their qualification, through publicly available information on the IFCN website. This information included the name of the fact-checking organisation, its location, and the language it publishes in. Some fact-checking organisations publish articles in multiple languages. In such cases, we accounted only for the primary language listed on the IFCN website \cite{ifcnsigs}. Our dataset of IFCN sources included 185 organisations from 70 countries writing in 48 different languages. Fifty-nine IFCN members write primarily in English, while 24 languages only have one IFCN-certified fact-checking organisation that writes primarily in that language. 

We began our analysis by parsing all URLs listed in the previous dataset of Community Notes gathered between September 2024 and June 2025. We then matched these URLs against the IFCN signatories' URLs. Finally, using X's data on helpfulness ratings, we determined the number of notes rated ``Helpful'', ``Not Helpful'', and ``Needs More Ratings'' across all languages and across IFCN-citing notes.

\subsubsection{Results}
We found that IFCN sources were only cited in 21,530 notes, as compared to over 649,828 notes without an IFCN source, representing 3.2\% of all Community Notes. However, in line with previous research \cite{maldita2025,borensteinetal2025}, we found that when cited, these notes have a higher success rate in terms of receiving enough positive ratings for publication (12.7\% of IFCN-citing notes were rated as Helpful versus 8.1\% non-IFCN notes rated as Helpful). They were likewise less likely to be rated as Not Helpful (1.2\% of IFCN-citing notes were rated as Not Helpful versus 3.8\% of non-IFCN notes). The vast majority of all notes, regardless of source, failed to receive enough votes to be rated as either Helpful or Not Helpful.
\begin{table}[h!]
\begin{center}
\begin{tabular}{|c|c|c|c|c|}
\hline
Language & \makecell{Number of Notes\\Citing IFCN Sources} & \makecell{IFCN Coverage} & \makecell{IFCN-Citing Notes \\Rated ``Helpful''} & \makecell{IFCN-Citing Notes\\ Rated ``Not Helpful'' }\\
\hline
English & 13,305 & 3.2\% & 13.75\% & 1.30\% \\
\hline
Spanish & 2,092 & 3.3\% & 10.80\% & 0.91\% \\ 
\hline
French & 1,841 & 5.3\% & 13.31\% & 1.03\% \\
\hline
Portuguese & 1,412 & 3.4\% & 12.32\% & 1.70\%\\
\hline
Japanese & 990 & 1.7\% & 25.15\% & 0.10\% \\
\hline
German & 519 & 3.5\% & 8.29\% & 0.58\% \\
\hline
Italian & 360 & 11.2\% & 6.94\% & 1.67\% \\
\hline
Turkish & 312 & 2.1\% & 25.96\% & 0.32\% \\
\hline
Polish & 142 & 2.6\% & 5.63\% & 0.00\% \\
\hline
Hebrew & 106 & 1.8\% & 17.92\% & 3.8\% \\
\hline
Indonesian & 95 & 8.3\% & 13.68\% & 0.00\% \\
\hline
Dutch & 77 & 3.8\% & 10.38\% & 0.00\% \\
\hline
Serbian & 53 & 2.9\% & 30.18\% & 0.00\% \\
\hline
Persian & 34 & 3.3\% & 17.64\% & 0.00\%  \\
\hline
Arabic & 32 & 1.4\% & 18.75\% & 0.00\% \\
\hline 
\end{tabular}
\caption{Languages with more than 1,000 total Community Notes and their citation of IFCN-certified fact-checking organisations. The ``IFCN Coverage'' column indicates the percentage of total notes in that language that cite IFCN sources.}
\label{table:2}
\end{center}
\end{table}

The top fifteen most common languages represented in Community Notes cited IFCN-certified sources to varying degrees. Arabic only cited IFCN sources 1.4\% of the time, whereas 11.2\% of Italian-language notes cited an IFCN source. Italian was not alone in having a relatively high degree of dependency on professional fact-checking sources. When looking at all languages in the dataset, there were eight languages in which more than 1 in 20 notes cited an IFCN source.  Some of these eight languages, however, have only a few notes in total, so it is hard to draw general conclusions about their dependency on IFCN sources. For example, Malayalam has only 34 notes in total, but 20\% of these cite an IFCN source. 

Among the thirty languages with more than 100 total notes, only 2 did not cite any IFCN sources (Danish and Urdu). It was surprising that Danish-language notes did not cite any IFCN sources as there is a Danish-language IFCN signatory, whereas there is no Urdu language IFCN source. Among those languages that have only one IFCN source writing in their language, Norwegian relied most heavily on IFCN sources, with 6.7\% of notes linking to an IFCN fact-checker, three-quarters of which linked to the Norwegian-language IFCN organisation.

While IFCN sources were cited in languages with fewer than 100 notes (2.6\% of notes from these languages cited an IFCN source), it is difficult to ascertain how helpful they were overall as there were so few notes written in each language and the vast majority failed to receive sufficient votes for classification as Helpful or Not Helpful. Thus, in assessing the impact of citing an IFCN source on helpfulness across languages, we narrowed our analysis to the ten languages which cite IFCN sources more than 100 times. Of these, in almost all languages that we analysed, notes that cited an IFCN source were as likely or more likely to be rated as helpful. Japanese had the largest difference, with only 11.55\% of notes overall rated as helpful, but 25.15\% of IFCN-citing notes rated as helpful. Turkish also had a large difference, with 16.5\% of notes overall rated as helpful, but 26\% of IFCN-citing notes rated as helpful. 

Italian is an interesting exception to this trend. Despite having the highest rate of citing IFCN sources among languages with more than 100 notes, IFCN-citing notes were not perceived as more helpful. While 7.97\% of Italian language notes overall were rated as Helpful, only 6.94\% of IFCN-citing notes were rated as helpful.  This could, perhaps, be due to the fact that more than three-quarters of Italian IFCN-citing notes linked to an IFCN source that writes primarily in a foreign language. However, Japanese also had high rates of citing foreign-language IFCN sources (55\% cited a foreign language source) and yet still had higher rates of helpfulness among notes that cited IFCN sources versus those that didn't. Furthermore, in most of the top ten languages, there was no obvious connection between the rate of citing a foreign-language source and the rate at which IFCN-citing notes were rated Helpful. 

We were, indeed, surprised at how often note-writers drew on fact-checking sources that do not write in the same language as the note. Excluding English-language notes, 30.6\% of notes that cite an IFCN source link to at least one that publishes primarily in a language different than that of the note. Sometimes, a note-writer will link to multiple IFCN sources, at least one writing in the same language as the note and at least one other writing in a different language. In some of these cases it is possible that the IFCN source publishes articles in several languages and thus the language of the note and the IFCN article are the same. However, most fact-checking sources do not publish in multiple languages, so it is unlikely that this is true of all cases. Of the top fifteen language languages, dependence on same-language fact-checking organisations varied considerably. As previously mentioned, while Italian-language notes had a high rate of citing IFCN sources overall (11.2\%), only 23.9\% of these cited an Italian-language IFCN source. Conversely, Indonesian-language notes also had a high rate of citing IFCN sources overall (8.3\%), 93.7\% of which cited Indonesian-language IFCN sources. Given that English is by far the most represented language among IFCN fact-checking organisations (59 out of 185 organisations publish primarily in English), it is unsurprising that over 90\% of English-language notes that cite an IFCN source cite one that is writing in English.  

\section{Discussion}
The above research suggests that it may take time for a Community Notes model of fact-checking to take hold in a given linguistic community or geographical area, even if there are a large number of active X users in that region. However, time and high levels of participation on X may not be sufficient to guarantee widespread participation in Community Notes. For example, Italy was added to the program at the same time as Germany (June 2023) and both, per Sohail et. al. 2025, are similarly active in producing tweets. Yet there are nearly five times as many German-language notes as Italian-language notes (14,891 notes to 3,224 notes). Similarly, X added Israel and a number of other Arabic speaking countries at the same time in November of 2023. Our data was collected almost a year later and only 2,269 notes in our dataset were written in Arabic, whereas there were twice as many notes written in Hebrew (5,816 notes), despite there being far more Arabic speakers and similar rates of participation on X in Israel and nearby Arabic speaking countries \cite{sohailetal2025}. 

This suggests that there are variables other than time, number of language speakers, and levels of participation on X that contribute to rates of volunteers writing fact-checks. As such, it is currently unclear whether the disparate rates of participation across linguistic communities represents a problem with Community Notes style programs in terms of their ability to manage misinformation. It is possible, for example, that some linguistic communities are less likely to share the kinds of empirical claim that are frequently the targets of fact-checking. If this were the case, then lower rates of participation would simply reflect lower rates of empirical misinformation being shared in that community. In linguistic communities where it is common to speak multiple languages, users may tend to write on X in one language more than another, thus leading to fewer notes in the less-used language. In such cases, the disparate rates of note-writing don't necessarily represent a challenge for managing online misinformation. 

Alternatively, however, some communities may have comparatively limited access to the informational resources necessary to perform fact-checking. As seen in Wikipedia's crowd-sourced knowledge work, omissions occur when there is a lack of written resources that are available via hyperlink. Similar omissions are likely to occur in crowd-sourced fact-checking because, like Wikipedia, Community Notes contributors are strongly encouraged to provide hyperlinks to relevant sources, and notes that lack reputable sources are less likely to be rated as Helpful \cite{solovevandprollochs2025}. Volunteer fact-checking is also a time-consuming and unpaid effort, which might prove a barrier for less economically well-off communities. Such communities may experience high rates of misinformation on X, but lack the social, material and information resources to address it themselves. Understanding what other variables contribute to disparate rates of engagement in writing Community Notes is thus vital for determining whether Community Notes is a suitable response to misinformation across all cultural and linguistic contexts. An important avenue for future research, then, is to understand what other variables contribute to disparate rates of participation across linguistic and geographical communities. This is particularly important if platforms rely solely on Community Notes-style programs for fact-checking.

Not only do linguistic communities need volunteers to write notes, they also need volunteers willing to rate these notes, which similarly requires time and informational resources. There is significant variation in the rate of notes reaching a status of Helpful and Not Helpful across languages. Having more notes written in a language does not guarantee higher rates of publication. To some extent, this should be expected as X's use of different algorithms across language groups is designed to increase the number of notes that reach a Helpful status in groups with fewer participants. However, we found that in the least represented linguistic groups, notes still fail to receive enough votes to reach a status of either Helpful or Not Helpful. The effectiveness of having a different algorithm to address lower rates of participation seems to bottom out as rates of participation drop below a certain threshold. Given our dataset, this seemed to be around 100 notes, as all languages with more than 100 notes had at least 1\% of notes rated as Helpful, whereas the majority of languages with less than 100 notes had none.

 
Given the overall low rate at which notes are rated Helpful, thereby appearing to the public, it is important to effectively leverage the informational resources that are available. It is clear that professional fact-checking sources represent a valuable resource for volunteer fact-checkers, although dependence on this resource differs across linguistic communities. In order to foster collaboration between professional and volunteer fact-checkers, it would be helpful to know why volunteers do not rely more heavily on their professional counterparts. Are professional fact-checkers too slow in producing fact-checks? Do volunteer fact-checkers not know about these resources or do they not trust them? Do volunteer fact-checkers address different kinds of content than professional fact-checkers, rendering the latter's work not applicable to their own? Future research should aim to identify whether and how professional and volunteer fact-checkers can work effectively with one another to manage misinformation online.
 
\subsection{Limitations}
There are a number of limitations with this study. The fastText language classification system offers different confidence thresholds for classification. We ran the analysis with a zero-percent confidence threshold, so that the model classified all notes, even ones that it may not have been confident about. The above discussion relates to the zero-confidence analysis. However, we also ran the analysis again with a 50\% confidence and 90\% confidence. When assigning notes to languages with these levels of confidence, a number of notes were left unclassified (19,564 unclassified notes at 50\% confident, 299,695 unclassified at 90\% confident). Thus, some of the notes in our discussion were likely misclassified. However, across all the confidence thresholds, there were no changes in which languages were in the top ten most represented languages and there were very few changes in the order of their representation. The top four languages at the 0\% confidence threshold were (1) English, (2) Spanish, (3) Japanese, (4) Portuguese. At 90\% confident, this shifted to (1) English, (2) Japanese, (3), Portuguese, (4) Spanish. These were the only changes in ordering in the top ten most common languages, indicating that the low confidence threshold captured the general patterns of note-writing frequency for the most represented languages. At 50\% confident, however, a number of the least represented languages fell out of the analysis, with only 8 languages listed with one note and at 90\% confident this number dropped to 5. Thus, our analysis is likely less accurate on the least represented languages.

A second limitation is that we did not have access to the number of tweets written in each language across the time period our data is gathered from. This made it difficult to assess how active various linguistic communities are on X outside of the Community Notes program. We relied on previous research that aimed to identify the geographic region of X users, but this is an imperfect proxy as not all languages are tied to geographic regions and that data is not drawn from the same time period. 

Additionally, this analysis is limited insofar as it focused on professional fact-checking organisations that are IFCN members, which disregards a large number of other organisations and media outlets that perform fact-checking, but are not IFCN signatories. This was intentional, as we wanted to assess, in part, whether speakers of languages with fewer professional fact-checking organisations would be negatively impacted if those organisations were to close without Meta's funding. However, given that many of the languages with fewer IFCN members also had relatively few notes written overall, we were unable to identify any dependencies between volunteer fact-checkers and these IFCN members.

Furthermore, we only accounted for the primary language that an IFCN member publishes in. Several of these organisations publish in multiple languages and thus, as noted earlier, the rate at which people cite notes in foreign languages may be slightly lower than reported here.

\section{Conclusion}
As platforms aim to identify effective measures to address misinformation, it is valuable to know whether these measures will work across different language groups. While some research has investigated the efficacy of crowd-sourced fact-checking in X's Community Notes program, these studies have focused on English-language notes or the program as a whole. In this study, we analysed rates of participation in X's Community Notes platform across languages in terms of how many users contribute notes per language, how many notes are written, how often they are rated Helpful and how often they cite professional fact-checking organisations. Our analysis implies that crowd-sourced fact-checking takes time to get established in some communities, but that time alone is not sufficient. In linguistic communities that have lower rates of participation, crowd-sourced fact-checking may be an ineffective measure for addressing misinformation. These findings have important implications for platforms regarding content moderation in smaller linguistic communities and in linguistic communities that have less social, material, and informational resources. These findings also mirror issues that exist in other cross-linguistic crowd-sourced knowledge production, in particular those of Wikipedia, wherein communities with fewer sources of written, publicly available information are less represented. As social media platforms increase their global presence, it is vital that they account for such linguistic communities when devising their content moderation strategies.

\section*{Data Availability Statement}
This research relied on publicly available and anonymous data from X's Community Notes program and did not need ethics approval. Data regarding IFCN sources is publicly available on their website (\url{https://www.poynter.org/ifcn/}).

\section*{Acknowledgements}
This research was funded by a grant from the Arts Digital Lab at the University of Canterbury (UCARTS-ADL-2025-01).

\section*{Declaration of Interest}
The authors report there are no competing interests to declare.

\section*{Author Contributions}
Elizabeth Stewart is the corresponding author and is responsible for the written analysis. Suryash Greenwold and Timotius Marselo conducted data collection and analysis.

\bibliography{Community_Notes_Bib}
\bibliographystyle{apalike}

\newpage
\section*{Table 1}
\begin{table}[h!]
\begin{center}
\begin{tabular}{|c|c|c|c|c|c|c|}
\hline
Language & Date Added & \makecell{Number\\of Contributors} & \makecell{Number\\ of Notes} & \makecell{Notes Rated\\ ``Helpful''} & \makecell{Notes Rated \\``Not Helpful'' }& \makecell{Notes Needing \\More Ratings}\\
\hline
English & Oct. 2022 & 104,852 & 413,950 & 7.81\% & 4.13\% & 88.06\% \\
\hline
Spanish & April 2023 & 18,802 & 62,753 & 9.15\% & 3.22\% & 87.63\% \\ 
\hline
Japanese & March 2023 & 11,494 & 58,776 & 11.55\% & 3.23\% & 85.22\% \\
\hline
Portuguese & March 2023 & 9382 & 42,140  & 12.95\% & 3.50\% & 83.55\% \\
\hline
French & Dec. 2022 & 9237 & 34,576 & 11.23\% & 4.10\% & 84.67\% \\
\hline
German & June 2023 & 4,440 & 14,891 & 7.14\% & 3.04\% & 89.82\% \\
\hline
Turkish & July 2023 & 3,819 &14,621 & 16.50\% & 3.05\% & 80.45\% \\
\hline
Hebrew & Nov. 2023 &1,361 & 5,816 & 12.00\% & 4.37\% & 83.63\% \\
\hline
Polish & July 2023 & 1,357 & 5,523 & 5.83\% & 1.53\% & 92.64\% \\
\hline
Italian & June 2023 &1,199 & 3,224 & 7.97\% & 3.26\% & 88.77\% \\
\hline
Arabic & Nov. 2023 & 896 & 2,269 & 7.27\% & 1.14\% & 91.32\% \\
\hline
Dutch & June 2023 & 782 & 2,013  & 6.86\% & 3.23\% & 89.91\% \\
\hline
Serbian & July 2023 & 494 &1,818  & 24.26\% & 0.99\% & 74.75\% \\
\hline
Indonesian & Nov. 2023 & 429 & 1,148 & 12.80\% & 1.13\% & 86.07\% \\
\hline
Persian & Sep. 2024 & 348 & 1,019 & 10.20\%& 2.06\% & 87.74\% \\
\hline
\end{tabular}
\caption{Languages with more than 1,000 Community Notes. The ``Date Added'' column details the date when Community Notes was first made available in a country where that language is an official or recognized language.}
\label{table:1}
\end{center}
\end{table}

\newpage
\section*{Table 2}
\begin{table}[h!]
\begin{center}
\begin{tabular}{|c|c|c|c|c|}
\hline
Language & \makecell{Number of Notes\\Citing IFCN Sources} & \makecell{IFCN Coverage} & \makecell{IFCN-Citing Notes \\Rated ``Helpful''} & \makecell{IFCN-Citing Notes\\ Rated ``Not Helpful'' }\\
\hline
English & 13,305 & 3.2\% & 13.75\% & 1.30\% \\
\hline
Spanish & 2,092 & 3.3\% & 10.80\% & 0.91\% \\ 
\hline
French & 1,841 & 5.3\% & 13.31\% & 1.03\% \\
\hline
Portuguese & 1,412 & 3.4\% & 12.32\% & 1.70\%\\
\hline
Japanese & 990 & 1.7\% & 25.15\% & 0.10\% \\
\hline
German & 519 & 3.5\% & 8.29\% & 0.58\% \\
\hline
Italian & 360 & 11.2\% & 6.94\% & 1.67\% \\
\hline
Turkish & 312 & 2.1\% & 25.96\% & 0.32\% \\
\hline
Polish & 142 & 2.6\% & 5.63\% & 0.00\% \\
\hline
Hebrew & 106 & 1.8\% & 17.92\% & 3.8\% \\
\hline
Indonesian & 95 & 8.3\% & 13.68\% & 0.00\% \\
\hline
Dutch & 77 & 3.8\% & 10.38\% & 0.00\% \\
\hline
Serbian & 53 & 2.9\% & 30.18\% & 0.00\% \\
\hline
Persian & 34 & 3.3\% & 17.64\% & 0.00\%  \\
\hline
Arabic & 32 & 1.4\% & 18.75\% & 0.00\% \\
\hline 
\end{tabular}
\caption{Languages with more than 1,000 total Community Notes and their citation of IFCN-certified fact-checking organisations. The ``IFCN Coverage'' column indicates the percentage of total notes in that language that cite IFCN sources.}
\label{table:2}
\end{center}
\end{table}

\end{document}